# Experimental research on the THGEM-based thermal neutron detector

YANG Lei(杨雷)[1]  Zhou Jian-Rong(周健荣)[2;3,*]  Sun Zhi-Jia(孙志嘉)[2;3]  Zhang Ying(张莹)[4]  Huang Chao-Qiang(黄朝强)[4]  Sun Guang-Ai(孙光爱)[4]  Wang Yan-Feng(王艳凤)[2;3]  Yang Gui-An(杨桂安)[2;3]  Xu Hong (许虹)[2;3]  Xie Yu-Guang(谢宇广)[2;3]  Chen Yuan-Bo(陈元柏)[2;3]

1 Dongguan University of Technology, Dongguan, 523808, China
[2] Institute of High Energy Physics, Chinese Academy of Sciences, Beijing 100049, China;
[3] State Key Laboratory of Particle Detection and Electronics,Beijing 100049,China;
[4] Institute of Nuclear Physics and Chemistry, China Academy of Engineering Physics, Mianyang, 621900,China

**Abstract:** A new thermal neutron detector with the domestically produced THGEM (THick Gas Electron Multiplier) was developed as an alternative to $^3$He to meet the needs of the next generation of neutron facilities. One type of Au-coated THGEM was designed specifically for the neutron detection. A detector prototype had been developed and the preliminary experimental tests were presented, including the performance of the Au-coated THGEM working in the Ar/$CO_2$ gas mixtures and the neutron imaging test with $^{252}$CF source, which would provide the reference of experimental data for the research in future.

## 1. Introduction

Neutrons are used to investigate the structure and dynamics of a material. Many efforts have recently been devoted to the development of the next generation of neutron facilities, which include SNS in USA, J- PARC in Japan, ISIS in UK, CSNS (China Spallation Neutron Source) in China and ESS in Europe [1]. The neutron detector is one of the key components of the neutron scattering instruments. With the international development of the new generation neutron source, the traditional neutron detector based on $^3$He has not been able to satisfy very well the demand of the application of high flux especially. And also facing the global crisis of $^3$He supply as well [2], the research on the new style of the neutron detector which can replace the $^3$He based detection technology becomes extremely urgent.

As a good candidate, a boron-coated GEM became the focus of attention recently [3,4], which was firstly designed by Martin Klein using CERN standard GEM in 2006[4]. It has the outstanding and excellent characteristics, such as high counting rate capability (>10MHz/mm$^2$), good spatial resolution and timing properties, radiation resistance, flexible detector shape and readout patterns[5]. In 2011, IHEP[6] and UCAS[7] firstly developed successfully a kind of THGEM, manufactured economically by standard printed-circuit drilling and etching technology in China. Compared with the CERN standard GEM, THGEM has higher gain, sub-millimeter spatial resolution and the possibility of industrial production capability of large-area robust detectors[8], which is very suitable for the application of neutron detection.

**Foundation item**:Supported by National Natural Science Foundation of China (Grant No.11127508, 11175199), NPL, CAEP (Grant No. 2013DB06) and the State Key Laboratory of particle Detection and Electronics (Grant No. H9294206TD)
**Corresponding author**: ZHOU Jian-rong,E-mail: zhoujr@ihep.ac.cn

In this paper, it would be presented the experimental research on a new kind of the neutron detector based on the domestically produced THGEM which was provided by Xie Yuguang group of IHEP. It's a Au-coated THGEM with a thickness of 300μm, hole diameter of 250μm, pitch of 600μm and a rim of 80μm. The THGEM was made of FR4 glass epoxy substrate, with copper cladding on both sides and then Au coated on the copper, which had an active area of 50mm*50mm. In order to study its basic characteristics as the references for the development of this kind of THGEM based neutron detector, the performances of the counting rate plateau, the energy resolution and the gain had been measured in the different $Ar/CO_2$ gas mixtures with the different high voltages. According to the tests, the working conditions optimized of the THGEM had been obtained.

By using this kind of THGEM, a detector prototype had been developed. In order to study the characteristics of the prototype including its data acquisition system and make a good preparation for the neutron beamline test in next step, a preliminary imaging measurement with $^{252}CF$ neutron source was carried out in IHEP. According to the experiments, it would be very helpful for the design of this kind of THGEM-based neutron detector in future[9,10].

## 2. Design of the detector prototype

The detector mainly consists of the neutron convertor, the single THGEM for gas multiplication(~100), and the 2-D strips structure for signal readout. Fig.1.a shows the schematic view of the detector and the Fig.1.b is the image of the detector. The resistance chain is used to provide three channels of high voltage supply. With a $^{10}B$ layer coated on the drift cathode as the neutron convertor, when a neutron is captured, either an α ion or a $^7Li$ ion is emitted into the drift gas volume, where it releases along its track a large number of Secondary Electrons (SE), which is proportional to the energy deposited in the drift gas volume (~$3\times10^4$ SE for 1 MeV). THGEM is employed for SE multiplication and can provide an effective gain of several hundreds. And then, the signal will be induced on the 2-D strips structure for determining the spatial and timing information by the readout electronics and data acquisition. The detector is operated in flow mode with $Ar/CO_2$ mixtures at atmospheric pressure. Continuous purge of cheap counting gas avoids ageing effects encountered in other detectors so as to get long term stability as well as long lifetime. Table 1 shows the design specifications of the detector prototype.

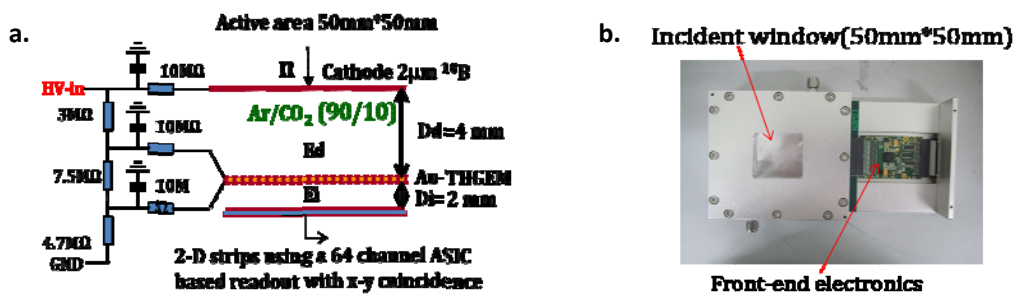

Fig.1. Schematic of the detector prototype

[键入文字]

Table 1. Specifications of the detector prototype

| PARAMETER | SPECIFICATION |
|---|---|
| Active Area | 50mm*50mm |
| Neutron Flux | $<10^8$n/cm$^2$.s |
| Spatial Resolution(FWHM) | <3mm |
| Timing Resolution | <1μs |
| Efficiency@1.8Å | ～4% |
| Max Counting Rate | >1MHz |
| Working mode | Real-time |

The data acquisition system of the detector is based on the analogue front-end readout ASIC chip and the modern Field- Programmable-Gate-Array (FPGA) technology(Fig.2). The 2-D strips is followed directly and read out by a front-end electronics of 64 channels CIPix ASIC based daughter board. The strip period is 1.56mm with a gap of 0.26mm. For each dimension, there are 32 channels for readout. The CIPix ASIC was originally developed by the Heidelberg ASIC lab in 2000 for the DESY H1experiment. A single chip integrates 64 channels of a low noise charge sensitive preamplifier followed with a shaper and discriminator. In the typical neutron detection, no external trigger is available. On the contrary, individual neutron events are entirely uncorrelated. The FPGA is employed to realize the high bandwidth data processing. As one neutron is detected, a cloud of electron charges, drifting onto the readout structure, will be collected by electrode strips corresponding to both dimensions x and y. Thus, the detection of a neutron at the corresponding (x,y)coordinates will be obtained by the means of simultaneous coincidence of signals on x- and y-readout channels. After coincidence, the time stamp of each neutron will be added and then an event reconstruction including the spatial and timing information(x,y,t) is completed. Finally, the data will be transferred to the PC for data acquisition and analysis.

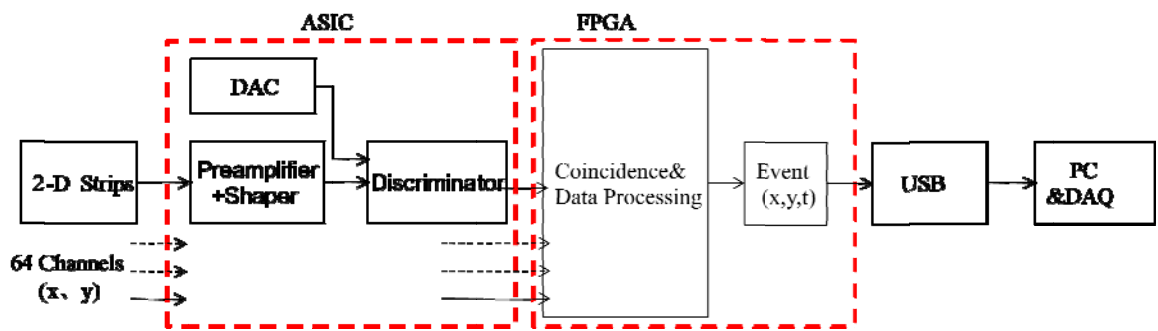

Fig. 2. Schematic of the data flow and processing

[键入文字]

### 3. Thermal neutron efficiency

Compared with the highly reactive and expensive $^6$Li, solid $^{10}$B layer seems to be much more favourable and suitable for use as neutron converter. It can easily be produced in reasonable sizes using evaporation or sputtering techniques. $^{10}$B (enrichment >99%) is commercially available. The enriched $^{10}$B is coated on one surface of the copper cathode plate. Neutrons are detected by the following neutron reactions:

$$n+^{10}B \rightarrow \alpha+^7Li + 2.79 MeV \quad 7\% \quad E_\alpha=1.78MeV \quad E_{Li}=1.0MeV$$

$$n+^{10}B \rightarrow \alpha+^7Li^* + 2.31 MeV \quad 93\%$$

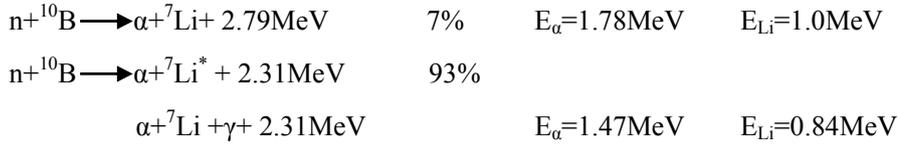

$$\alpha+^7Li + \gamma + 2.31 MeV \quad E_\alpha=1.47MeV \quad E_{Li}=0.84MeV$$

When the thermal neutron induces the $^{10}$B(n, α)$^7$Li reaction, 93 % of all the reactions lead to the first excited state of $^7$Li$^*$, which decays spontaneously (~73fs half-time) to the ground state of $^7$Li by emitting the 0.48 MeV gamma ray, and 7% of the reactions result in the ground state of $^7$Li. The thermal neutron (0.0253eV) cross section of the $^{10}$B(n, α)$^7$Li reaction is 3840 barn and it drops rapidly with increasing the thermal neutron energy. The maximum conversion efficiency is about 5% for thermal neutrons through a single pure $^{10}$B-layer with the thickness of 2.5μm. For charged ions α and $^7$Li, the gas detector can reach its efficiency very close to 100%. As a result, the neutron efficiency can be considered as the conversion efficiency. For the thickness of 2.0 μm used in the prototype, the thermal neutron efficiency calculated by Geant4 is 4.6%.

### 4. Results and discussions

### 4.1. Tests on THGEM foil

In order to study its basic characteristics as the references for the development of this kind of THGEM based neutron detector, the performances of the counting rate plateau, the energy resolution and the gain had been measured in the different Ar/$CO_2$ mixtures with the different high voltages. The THGEM foil was tested by using a $^{55}$Fe X ray source (activity 10mCi) which was positioned in such a way that a collimated beam (φ1 hole) of X-rays perpendicularly entered the upper drift region. The signal was induced by a cathode pad and read out with an ORTEC 142IH preamplifier followed by an ORTEC 572A amplifier (shaping time t=2μs) and an ORTEC multi-channel analyzer (trump-usb-8k).

First of all, several THGEM foils were tested to check the stability of the gain in 12 hours. And then one of this foils which showed the best performance was chosen for the next tests below. After 200 minutes, the effective gain would get stable as shown in the Fig.3.a. Consequently, the THGEM warmed up about 200 minutes before every test. Fig.3.b shows the gain in the different Ar/$CO_2$ mixtures with the different high voltages. The THGEM can provide an effective gain range from 300 to 3000 in Ar/$CO_2$ mixtures. For the gas mixture Ar/$CO_2$ (90%/10%), the working voltage will be lower to give the same gain.

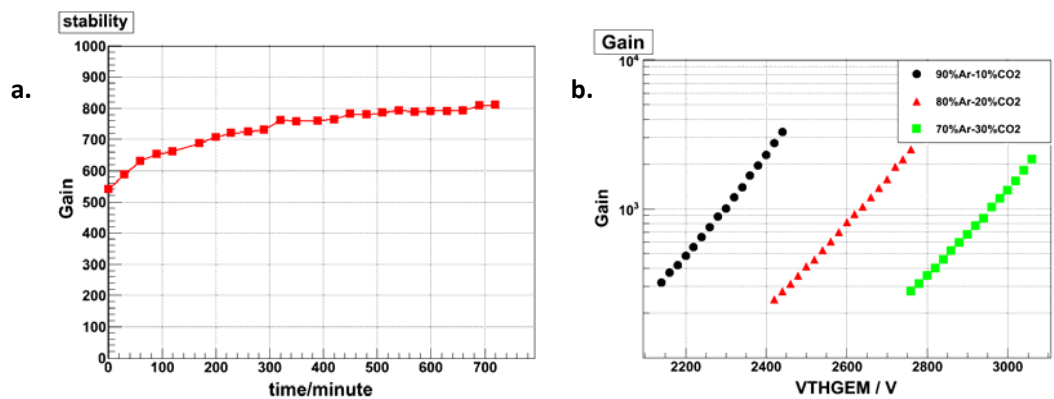

Fig.3. Gain and its stability of the THGEM

In order to know the suitable working voltage of the detector in different $Ar/CO_2$ mixture, its counter plateau was measured in different total voltage. During the experiment, the total flow of Ar and $CO_2$ gas is 50 SCCM to ensure the amount of effective working gas in the chamber. The counts were recorded in every one minute and the voltage of THGEM was increased by the increment of 20V until the spark discharge occurred. As the Fig.4 shows, it has a longer plateau in the $Ar/CO_2$ mixture ratio of 90%/10% and the THGEM woks in a lower voltage. For the gas mixture $Ar/CO_2$ (90%/10%), the plateau range of the THGEM is from 2200V to 2400 V and its plateau slope is smaller than 3% / 100 V, as shown in the left of the Fig.4. This optimization would be helpful to know the working range of the THGEM and find the conditions at the lower HV.

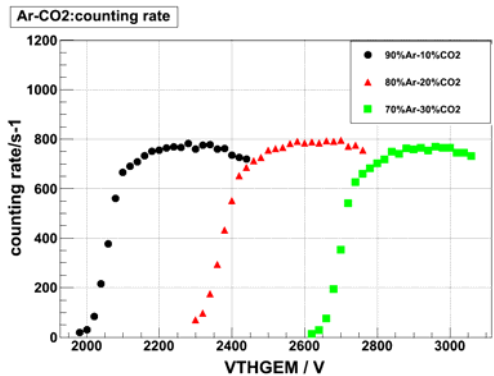

Fig.4. Counting rate plateau with the total HV

Energy resolution is one of the most important parameters related to the detector performance. By using the 5.9 keV $^{55}Fe$ X-ray source, the energy resolution was measured. The much better energy resolution is obtained in the $Ar/CO_2$ (90%/10%) gas mixture. Fig.5 shows the pulse height spectrum obtained with a $^{55}Fe$ source in the $Ar/CO_2$ (90%/10%) gas mixture. For obtaining energy resolution, it's fitted with the Gaussian function. It indicates the energy resolution (FWHM) of the detector based on THGEM is about 34% at 2440V. With such an energy resolution, the detector can entirely separate the 3keV of Ar escape peak from the $^{55}Fe$ main X-ray peak located at 5.9keV.

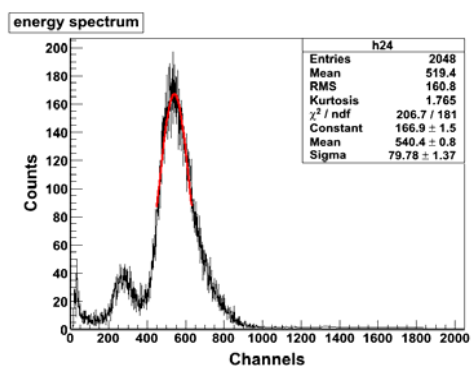

Fig.5. A pulse height spectrum in the Ar/$CO_2$ (90%/10%) gas mixture

### 4.2. Neutron tests in $^{252}$CF source

In order to study the characteristics of the prototype including its data acquisition system and make a good preparation for the neutron beamline test in next step. In IHEP, a preliminary measurement was carried out with an isotopic neutron source $^{252}$CF(~$1\times10^6$n/s), which was placed in a special shielding box with an open hole with a diameter of 100 mm in the side. The detector was mounted onto the exit of the hole. For obtaining larger counts, no neutron moderation was used in the test.

The neutrons emitted from $^{252}$CF source follow the fission spectrum and its average energy is 2.3MeV. The single cadmium mask cannot absorb all the neutrons. For the measurement of the spatial resolution, the mask with a slit of 3mm wide was made of 25mm thick boron plastic and 1mm thick cadmium. Due to the weak source, the test took 24 hours. The 2-D counts image is shown in the Fig.6.a and Fig.6.b is the projection on y axis. The counts distribution is fitted with the Gaussian function and the FWHM of spatial distribution is 4.2mm, which is a bit worse. The reason is because the neutrons emitted from $^{252}$CF source have lots of energy and the mask with the slit cannot absorb all the neutrons completely. This is the main reason why a monoenergetic neutron beam line is needed for the accurate measurement.

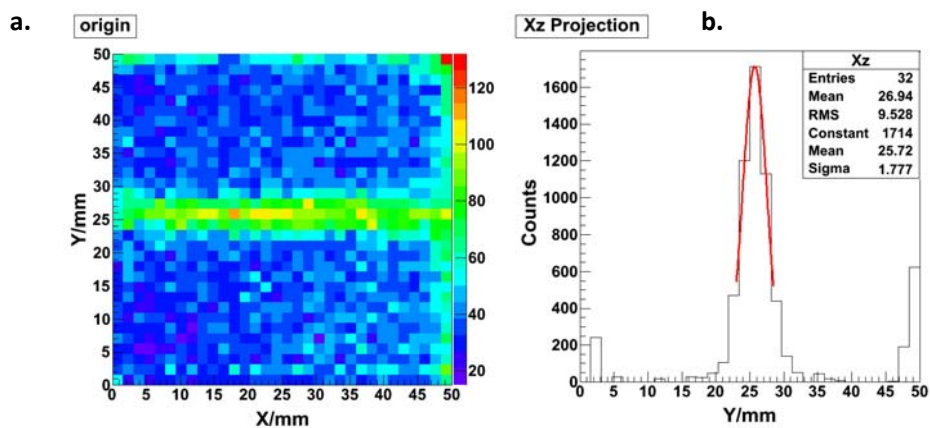

Fig.6. Spatial distribution

For testing the imaging ability of the whole system, the mask with a letter E was used, which was made of 25mm thick boron plastic and 1mm thick cadmium. The test took 24 hours as well. Due to the same reason, the image (Fig.7) is not good.

[键入文字]

However, it is sure that the detector prototype works normally as expected for neutron detection.

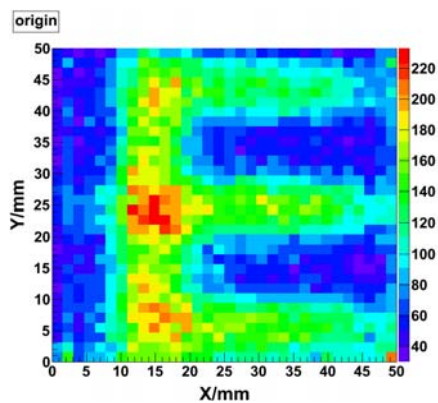

Fig.7. Neutron imaging

## 5. Summary

In this paper, a new type of neutron detector based on domestically produced Au-coated THGEM is introduced and the preliminary experimental study is presented, including the performance of the Au-coated THGEM working in the Ar/$CO_2$ mixtures and the neutron test of the detector prototype using $^{252}$CF source. This work is considered as the preparation of the accuracy measurement at the neutron beamline in the reactor, which will be reported in the near future.

## Acknowledgements

We are grateful to the supports from China Spallation Neutron Source, the State Key Laboratory of Particle Detection and Electronics, the Laboratory of Neutron Detection and Fast Electronics Technology in Dongguan University of Technology, the Laboratory of Neutron Physics in Institute of Nuclear Physics and Chemistry, China Academy of Engineering Physics.

[键入文字]

[键入文字]